\documentstyle[amsfonts,epsfig]{aipproc}

\def\NPB{{\em Nucl. Phys.} B}
\def\PLB{{\em Phys. Lett.}  B}
\def\PRL{{\em Phys. Rev. Lett.}}
\def\PRD{{\em Phys. Rev.} D}
\def\ZPC{{\em Z. Phys.} C}

\def\ra{\rightarrow}

\def\tb{\tan \beta}

\def\be{\begin{equation}}
\def\ee{\end{equation}}
\def\bea{\begin{eqnarray}}
\def\eea{\end{eqnarray}}

\def\gsim{\:\raisebox{-0.5ex}{$\stackrel{\textstyle>}{\sim}$}\:}

\def\eg{ {\it e.g. }}
\def\ie{ {\it i.e. }}
\def\bi{\bibitem}
\begin{document}
\begin{flushright}
IFT 98/4\\[1.5ex]
{\large \bf hep-ph/9803484}\\
\end{flushright}
\vskip -0.5cm
\title{Testing  2HDM at  Muon Colliders}
\author{Maria Krawczyk
\thanks{
The talk given at the 
Workshop on Physics at the First Muon
Collider and at the Front End of a Muon
Collider, FERMILAB, November 6-9, 1997.
Supported in part by
US-Poland Maria Sklodowska-Curie Joint Fund II (MEN/DOE-96-264).}}
\address{Institute of Theoretical Physics, Warsaw University, Warsaw, Poland}
\maketitle
\begin{abstract}
One very light neutral Higgs scalar with mass, say,  below 40-50 GeV   
is still allowed in the non-supersymmetric  2HDM (Model II),  
while the remaining particles of the Higgs sector in this  model
have to be heavier. 
The possibility of testing such a scenario at Muon Colliders
is discussed.
\end{abstract}
\section*{Introduction}
The   Two Higgs Doublet extension 
of the Standard Model (SM)  contains five Higgs particles: 
two neutral scalars $h$, $H$ ($M_H>M_h$), one pseudoscalar $A$
and two charged particles $H_{\pm}$, and  is characterized by 
two parameters $\alpha$ and $\tb$ 
(for  the CP conserved approach) , see \cite{hunter}.
Within this extension we study the non-supersymmetric version of the 
Model II, called here  2HDM. In this framework,
unlike in  the MSSM,  \footnote{Note  that    the  Higgs sector of the MSSM 
also belongs  to the  Model II.} all the  masses and $\alpha$ and $\tb$
are independent parameters, which  should be constrained one by one.

 The present 
mass limit on the Higgs scalar in the SM is 77 GeV \cite{smhiggs}.
The non-minimal  MSSM neutral Higgs bosons $h$ and $A$, 
for the considered (as a) typical 
supersymmetric particle spectra, individually  
should be heavier than $\sim$ 60 GeV (for $\tb >$ 1) \cite{mssm}. 
In contrast to the above limits 
the low mass range of the neutral Higgs sector
is  still allowed in  the non-supersymmetric case, \ie in  2HDM. 
In particular,
one very light neutral Higgs particle $h$ or $A$ may exist, 
even  as  light as 10 to 20 GeV
yet with the $\tb$ quite large, $\tb \sim$ 25 to 45 \cite{mk}
\footnote{It may be even lighter for  the appropriate
limit on the allowed range of  $\tb$ \cite{wilk,yukawa,mk}.}.
Such an object should 	in principle lead to visible effects.
 In fact, however, such a scenario is not yet excluded,
since some of the  effects are surprisingly   weak.

One should keep in mind that in 2HDM the mass limits for the lightest
neutral  Higgs bosons are of the form: $M_h$+$M_A\gsim $90-110 GeV,
therefore there is still a possibility
that both of these particles are relatively light, with mass well below
$M_Z$. Also it is worth   noticing  here that a
 very large mass gap between the
neutral Higgs bosons may naturally occur in 2HDM
 \footnote{In MSSM  
the  degeneracy in $M_h$ and $M_A$ is expected for large $\tb$.}.

As far as the charged Higgs boson is concerned it 
 should be at least as heavy as 330 GeV\cite{bsg,mi}, 
leading in most cases to negligible contributions to the observables.

The detailed study on the Higgs boson search at Muon Colliders both for the 
SM and  MSSM, can be found in proceedings  of the Muon Colliders workshops
and of this workshop, devoted to the physics at the First Muon Collider
\cite{muon}.
The aim of my talk is to present   unique features of the 2HDM
with a possibly one (very) light neutral Higgs boson. 
\section*{Status of the 2HDM with a light Higgs boson}
We start  the discussion of the 2HDM with the presentation of
 widths expected for the  lightest neutral Higgs bosons $h$ and $A$.
These widths   depend
crucially on the value of $\tb$, as $h$ and $A$ couple to fermions
by the coupling $\sim$ $(\cos(\alpha)/\sin(\beta))^{\pm}$ and 
$\tb^{\pm}$ for the states with $I_3$=(1/2)$^{\mp}$. 
  In the Fig.1 the LO results are presented
  for  the mass range below 40 GeV, where for  simplicity 
  we assume  $\alpha=\beta$.
\begin{figure}[hb]
\vskip 3.3in \relax\noindent\hskip 0.5in
                \relax{\includegraphics{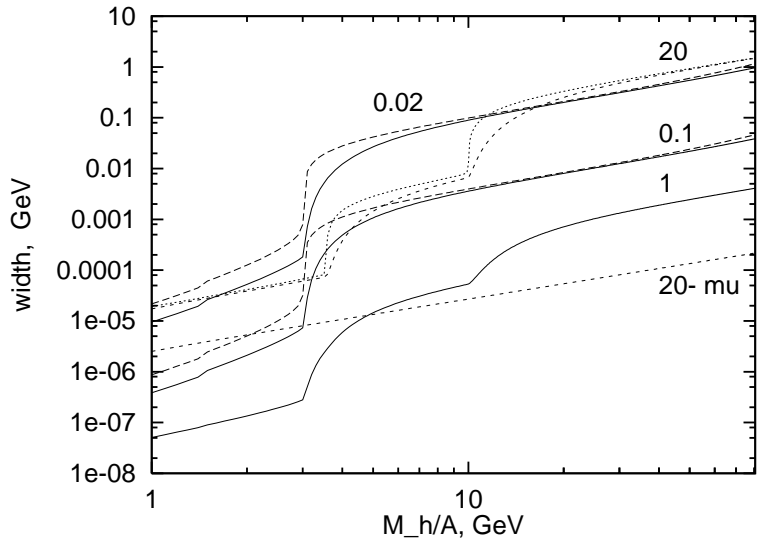}}
\vspace{-6.4ex}
\baselineskip 0.4cm
\caption{\sl  The width of  the scalar Higgs boson $h$
with $\alpha=\beta$ (solid line) and  of the pseudoscalar Higgs boson $A$
(dashed line) for  $\tb=0.1,\rm ~and ~0.02$. The results for $\tb$=20
are denoted by  short dashed and dotted lines for h and A, respectively.
In addition the SM prediction 
($\tb=1$) and the contribution from the 
muonic channel for $\tb=20$ (one line for 
both $h$ and $A$) are given.}
\end{figure}
The width of the SM Higgs scalar  (\ie for $\tb$=1)
is small in the whole considered mass region. 
For a large $\tb$ the width is considerably larger, reaching for $\tb$=20
and Higgs boson mass 40 GeV the value of 1 GeV.
It does not mean that for  small $\tb$ the width is always small,
see   Fig.1 for results for $\tb$= 0.1 and 0.02.
Note that, for the mass range 3-10 GeV,
 the widths obtained for $\tb$=20 and 0.1
 coincide. Similar effect can be found for 
the mass above 10 GeV, where the predictions for $\tb$=20 and 0.02
are close to each other.

Similarly to the widths also
the preferred decay modes for the studied by us light 
neutral Higgs bosons strongly 
 depend on the parameters of the model, mainly on   $\tb$.
Branching ratios of $h$ and of  $A$  can be found in Figs. 2a and 2b,
where  decays of the scalar and pseudoscalar are   presented
for  two  choices of  $\tb$,  $\tb=0.1$  
and $\tb=20$, respectively (also here 
for the simplicity we assume that $\alpha=\beta$).
\begin{figure}[ht]
\vskip 2.8in \relax\noindent\hskip -1.in
                \relax{\includegraphics{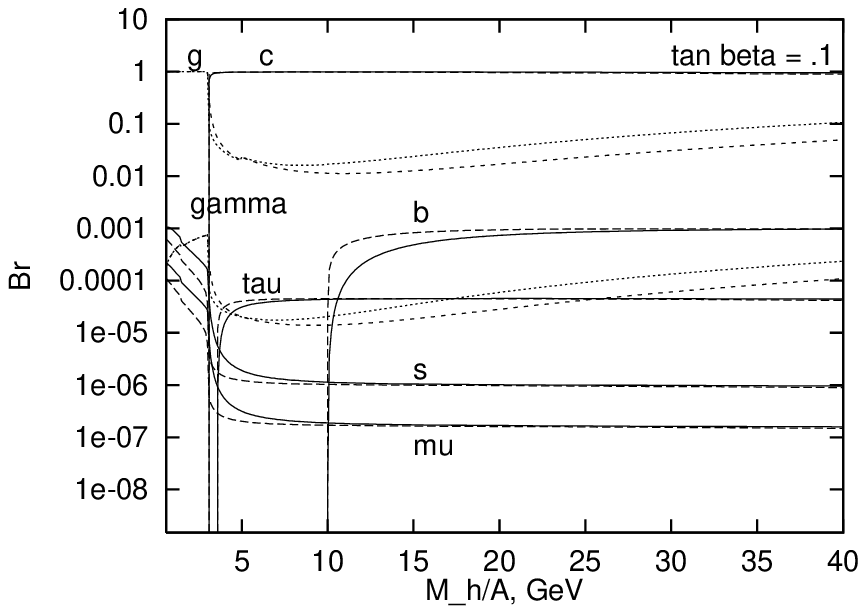}}
                  \relax\noindent\hskip 3.3in
                \relax{\includegraphics{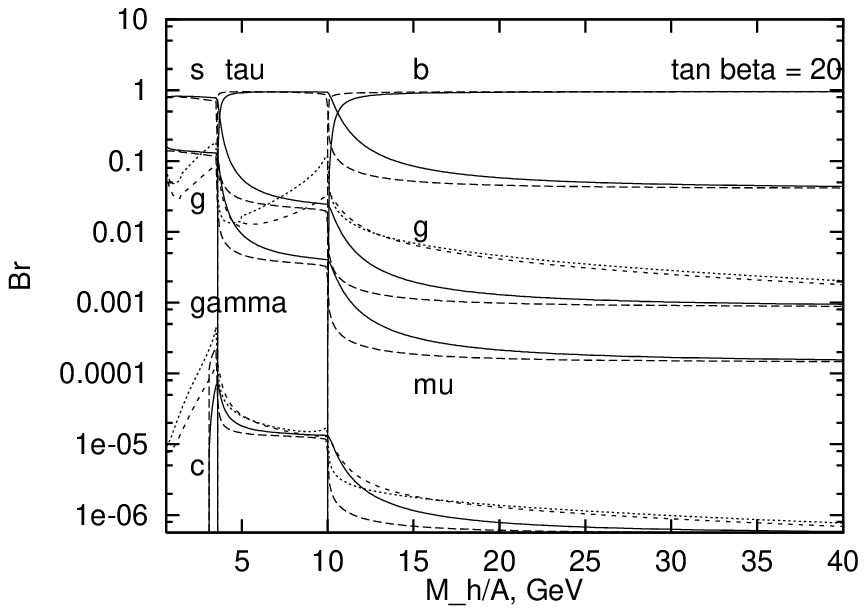}}
\vspace{-5.5ex}
\baselineskip 0.4cm
\caption{\sl  The branching ratios for   the scalar  $h$
with $\alpha=\beta$ (solid and short dashed line) 
and  for the pseudoscalar $A$
(dashed and dotted line) for a) $\tb=0.1$, b) $\tb=20$.}
\end{figure}

The basic  LEP limits on the parameters of 
the considered 2HDM model may be listed  as follows:
\begin{figure}[hc]
\vskip 3.5in \relax\noindent\hskip 0.in
                \relax{\includegraphics{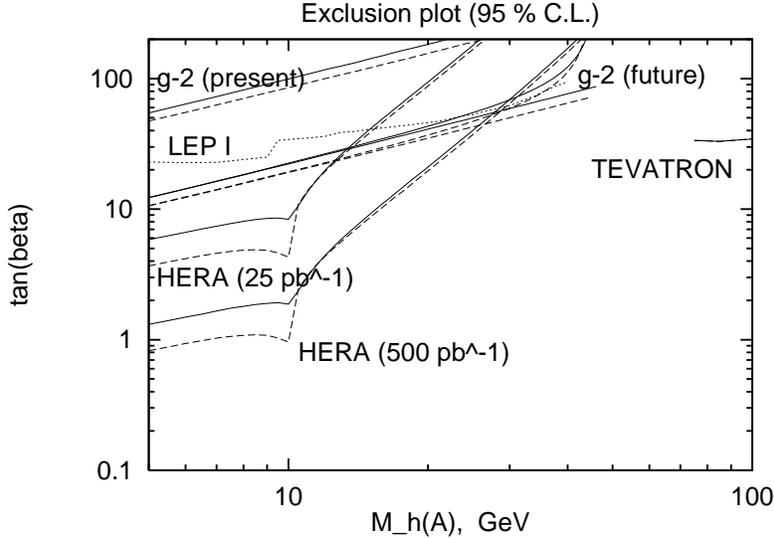}}
\vspace{-10.5ex}
\baselineskip 0.4cm
\caption{\sl The exclusion plot based on $g-2$ measurements, LEP data 
on the Yukawa process and TEVATRON data on $bb\tau\tau$, 
and on possible  results from the $ep$ collider HERA
 and from a low energy NL collider. Solid lines correspond to 
$h$ ($\alpha=\beta$), dashed and dotted lines to $A$. }
\end{figure}
\begin{itemize}
\item the 95\% C.L. exclusion derived from the Bjorken process,
$e^+e^-{{\ra}} Z h$,
on ~$\sin^2(\alpha-\beta)$ for $M_h$ smaller than 70 GeV \cite{sin},
indicates the small  value of  $\sin^2(\alpha-\beta)$, \eg
below 0.1 for $M_h\le$ 50GeV.
\item  one neutral Higgs boson can be very light, as 
the excluded region 
from the combined results on     $\sin^2(\alpha-\beta)$ and 
$\cos^2(\alpha-\beta)$  
(from the Bjorken process and the  Higgs boson pair production 
$e^+e^-{{\ra}} h A$,
 respectively) has a form $M_h+M_A < $ 90-110 GeV \cite{mhma}.
\item the 95\% C.L limit for $\tb$ for the $M_A$ smallar then 40 GeV 
\cite{yukawa}, obtained 
from the Yukawa process $e^+e^-{{\ra}} f {\bar f}   h/A$, 
 allows for   $\tb$ up to  20-25 for 
$M_A \sim 5-10 $ GeV and up to  100 for $M_A \sim 40$ GeV.
\item measurements of the process $e^+e^- {\stackrel {Z}{\ra}} h/A \gamma$
for $\tau \tau$, light quarks, $b$-quarks decay channels
performed recently at LEP collider by all experimental  groups 
\cite{zha} lead to relatively weak limits on  $\tb$ in 2HDM: $\tb\gsim$ 0.12
for $M_A\sim $10 GeV,  
as discussed  in Ref.\cite{kmz}.
\end{itemize} 
In addition, as it was mentioned above,  
$M_{H_{\pm}}$ should be larger than 330-350 GeV
as follows from the $b\ra s \gamma$ data\cite{bsg}
 and the newest NLO calculation
\cite{mi}. 
Note that the above limits from LEP, 
especially from the measurement of the Yukawa process 
and the $Z\ra h/A \gamma$ decay, may be much more stringent if the
luminosity will  be higher by factor 10 or so.

 The  present limits on $\tb$ in 2HDM from  LEP and TEVATRON \cite{drees}, 
$g-2$ measurement for the muon,  
together with the potential of the 
improved measurement of $g-2$ for the muon in the E821 experiment at BNL,  
and of  the HERA collider as well as  the NL collider 
\cite{mk} can be found in Fig.3.
\section*{Search for a light neutral Higgs boson at Muon Colliders}
The present constraints on  Higgs boson 
masses (and other parameters) of the general 2HDM
may be used to study   the  
potential of the Muon Colliders for searching for
light neutral Higgs bosons, 
if the tight constraints will not appear earlier.

We will concentrate below on the  potential of a direct search
of neutral Higgs bosons
at the First Muon Collider (FMC).  The energy of a collision 
at FMC is planned to be around the mass of the Z resonance, 
and in the next stage around 300-500 GeV, with 
the expected luminosity   1-10 fb$^{-1}$/yr.
The end-front search will not be  considered here although 
it may be very useful in the  search of a light Higgs particle in 2HDM.

From the point of view of the potential of First Muon Collider 
one has to distinguish two different Higgs mass ranges allowed in 2HDM:
\begin{itemize}
\item {\it the mass of the lightest Higgs boson is below $M_Z$}\\
This mass range  is unique for the considered by us 2HDM,
being excluded in the SM and in the standard 
scenarios discussed in the MSSM  (SUSY) approaches. 
In such case the study of the Higgs sector similar 
as at LEP I collider  can be performed, so 
obviously there is a need of  a higher luminosity than the one
obtained at LEP I.

Let me first concentrate on the mass of the lightest boson 
{\it below 40-50 GeV}.
In this case the analogous production mechanisms as discussed above 
for LEP I, with a dominating   $Z$- boson intermediate state, are possible -
in particular the Bjorken process, the Higgs pair production 
and the Yukawa process.

In the case of the production of the $h/A ~+ ~photon$ final state
 an important difference is expected in  comparison to the  LEP I
 measurements. 
In  $\mu^+\mu^-$ collision 
not only the loop but also the tree diagram contributes, 
being negligible for $e^+e^-$ collider.
Therefore in addition to the $Z$ pole processes 
the resonance production of the light neutral Higgs $h/A$
is also possible due the {\it return to the pole } (tree) process:
$\mu^+\mu^-\ra h/A \gamma$ \cite{tao,mu}. 

Note that in principle even {\it two} light Higgs bosons $h$ and $A$ 
with masses {\it around 40-50 GeV} may be produced at First Muon Collider.

If the mass of $h$ or $A$ is {\it close to $M_Z$} the scan in the energy  
(the mass) planned at FMC will allow to discover the resonant object. 
For large $\tb$ case the width of a light neutral  Higgs boson
is not small, also a  small   $\tb$ (as small as 0.1) may lead to
relatively large width (see Fig.1). So no extra requirement
referring to the beam energy resolution is needed for such a search.

The interesting option with two lightest neutral Higgs bosons $h$ and $A$
  in the direct reach of the FMC, still being very different 
in masses ({\it a large mass gap}),  is also open in the framework of 2HDM. 
In this case the corresponding processes as listed in the previous section 
for the LEP I collider are possible, with the $Z$ to be interchanged 
by the  heavier of two bosons.  A very attractive 
possibility of having as a resonance the 
heavier Higgs boson with the  lighter one in the final state  may occur, 
\eg  for $M_A\ge M_h$ the process 
$\mu^+ \mu^-{\stackrel {A}{\ra}} f {\bar f} h$.
\item {\it the mass of the lightest Higgs boson is above $M_Z$ but below 
the energy of the collider} \\
This case may have a similar signature to  the 
MSSM Higgs bosons production.
The basic difference to MSSM, beside the  fact that  $\alpha$ and $\beta$
parameters are not correlated with  the masses of Higgs bosons,
 is the possibility 
to have {\it a large mass gap} between two lightest neutral Higgs particles
$h$ and $A$.
Both particles may in principle be found as  resonances,
 the difference to the SM Higgs scalar case 
 is that their  widths may be large.
\end{itemize}

To summarize, the general 2HDM may lead to unique signature both
in the resonant production of neutral Higgs bosons, 
and  in the {\it higgsstrahlung} processes at
low energy First Muon Collider. Useful option with $\sqrt s \sim M_Z$
requires the luminosity at least by the factor 10 higher than the 
one achieved at LEP.
\vskip 0.2cm
\noindent
{\bf{\large{Acknowledgments}}}\\
The author is  very grateful 
to the Organizers of the Workshop for the invitation to this
interesting workshop and many important discussions. 
She also  likes to thank for the financial support.
     
\end{document}